\documentclass[conference]{IEEEtran}
\IEEEoverridecommandlockouts
\usepackage{cite}
\usepackage{amsmath,amssymb,amsfonts}
\usepackage{algorithmic}
\usepackage{booktabs}
\usepackage[OT1]{fontenc} 
\usepackage{graphicx}
\usepackage{subfigure}
\usepackage{textcomp}
\usepackage[misc]{ifsym}
\usepackage{xcolor}
\def\BibTeX{{\rm B\kern-.05em{\sc i\kern-.025em b}\kern-.08em
    T\kern-.1667em\lower.7ex\hbox{E}\kern-.125emX}}
\begin{document}
\title{iWave3D: End-to-end  Brain Image Compression with Trainable 3-D Wavelet Transform}
\author{Dongmei Xue, Haichuan Ma, Li Li${^\textrm{\Letter}}$, Dong Liu, Zhiwei Xiong \\
\textit{University of Science and Technology of China, Hefei, China} \\
\textit{\{xdm1, hcma\}@mail.ustc.edu.cn, \{lil1, dongeliu, zwxiong\}@ustc.edu.cn}}

\maketitle

\begin{abstract}
With the rapid development of whole brain imaging technology, a large number of brain images have been produced, which puts forward a great demand for efficient brain image compression methods. At present, the most commonly used compression methods are all based on 3-D wavelet transform, such as JP3D. However, traditional 3-D wavelet transforms are designed manually with certain assumptions on the signal, but brain images are not as ideal as assumed. What's more, they are not directly optimized for compression task. In order to solve these problems, we propose a trainable 3-D wavelet transform based on the lifting scheme, in which the predict and update steps are replaced by 3-D convolutional neural networks. Then the proposed transform is embedded into an end-to-end compression scheme called iWave3D, which is trained with a large amount of brain images to directly minimize the rate-distortion loss. Experimental results demonstrate that our method outperforms JP3D significantly by 2.012 dB in terms of average BD-PSNR. 

\end{abstract}

\section{Introduction}
In recent years, the rapid development of brain imaging technology has promoted the advancement of many fields such as medicine and artificial intelligence. However, as the center of the nervous system, the brain usually has a complex structure, and imaging it will generate massive amounts of data.
For example, the brain of drosophila contains about $10^{5}$ of neurons, and its raw images occupy about \(106\) terabyte (TB) of storage space\cite{zheng2018complete}. The human brain has more than $1.5\times10^{10}$ neurons, and the space required to store its raw images will be unimaginable.
This puts forward a great demand for efficient brain image compression methods.

A series of image compression methods have been proposed to solve this problem. JP3D\cite{bruylants2009jp3d} is the most commonly used.
JP3D was proposed as a compression standard to support the compression of 3-D images. It is a 3-D mode extension of JPEG-2000\cite{rabbani2002overview}. The building blocks of JP3D include preprocessing, discrete wavelet transform (DWT), quantization and entropy coding. Among them, wavelet transform is recognized as a powerful tool for time-frequency analysis. It is also widely used in other compression methods\cite{sathiyanathan2018medical,ravichandran2016performance,sriraam20113,sidhik2015comparative,boujelbene2019comparative,wang1996medical,bruylants2015wavelet,selvi2017ct}.

However, traditional wavelet transform still has some drawbacks. Firstly, the traditional wavelet transform is essentially 1-D. To perform a 3-D transform, it usually needs to be used in sequence in the axial, horizontal and vertical directions. Thus a large number of non-directional features cannot be captured. What's more, the traditional wavelet transform utilizes the same kernel on the entire image, and cannot handle different local features such as edges and textures. Additionally, the parameters of wavelet transform need to be manually designed to fit different data.
Last but not least, the traditional wavelet transform is fixed and cannot be optimized for compression tasks. 

In order to solve these problems of traditional wavelet transform, we propose a trainable wavelet transform based on 3-D lifting scheme\cite{calderbank1998wavelet,sweldens1998lifting,daubechies1998factoring,sweldens1995lifting,blessie2011image,sweldens1996lifting}, in which the predict and update steps are replaced by 3-D convolutional neural networks. Unlike the 1-D filter of the traditional wavelet transform, the 3-D convolution kernel can capture the features both directional and non-directional. In addition, it is reported that different convolution kernels in neural network usually focus on different parts of the image, such as texture and edges\cite{dosovitskiy2016inverting,mahendran2015understanding,yosinski2015understanding}. To this end, rich local features can also be handled. Moreover, the appropriate parameters of our transform are learned through a large number of training data and manual parameter selection is avoided. This means that our transform is data-driven and not only designed for a specific signal. Finally, we adopt an entropy coding module and a post-processing module to achieve an end-to-end compression framework called iWave3D. All the modules are jointly trained with a large number of brain images to minimize the rate-distortion loss. It is worth mentioning that Ma et al.\cite{ma2019iwave,ma2020end} proposed a trainable wavelet transform called iWave for 2-D image compression but they only studied on natural images. Differently, in this paper, we embed the 3-D trainable wavelet transform in an end-to-end compression framework named iWave3D for 3-D brain images compression.

We also find that our method is more suitable to compress the brain image compared with other popular end-to-end 2-D image compression methods based on auto-encoder.
Recently, deep learning-based end-to-end image compression has made great progress\cite{balle2016end,balle2018variational,minnen2018joint}. The performance reported in \cite{minnen2018joint} has even surpassed the most advanced traditional image compression method BPG.
Then is it feasible to directly transfer these methods to brain image compression? The answer is negative. It is reported in \cite{helminger2020lossy} that most of the end-to-end compression methods have poor performance at high bit rates. Their transform modules are composed of simply stacking convolutional layers, which lose too much information when converting images into latent features.
Unfortunately, high-quality reconstruction is often required to avoid fatal mistakes in brain image compression area, and some application scenarios even need lossless reconstruction. 
Unlike these methods, the proposed 3-D trainable wavelet transform is reversible, thus our method can reconstruct brain images with high quality, or even with no error.

Our main contributions can be summarized as follows:

\begin{itemize}
\item  We propose a 3-D trainable wavelet transform, which solves the problems of traditional wavelet transforms that are designed manually with certain assumptions on the signal and cannot be optimized for compression task. 
\item By cooperating the proposed transform with a entropy module and post-processing module, we form an end-to-end 3-D image compression framework named iWave3D.
\item Our iWave3D outperforms JP3D by a large margin in terms of PSNR and SSIM.
\end{itemize}

This paper is organized as follows. In section \ref{2}, we introduce the proposed iWave3D in detail. After that, section \ref{3} gives the detailed experimental results. Finally, the last section concludes the whole paper.

\section{Proposed Method}
\label{2}
\begin{figure}[t]
\centerline{\includegraphics[width=\columnwidth]{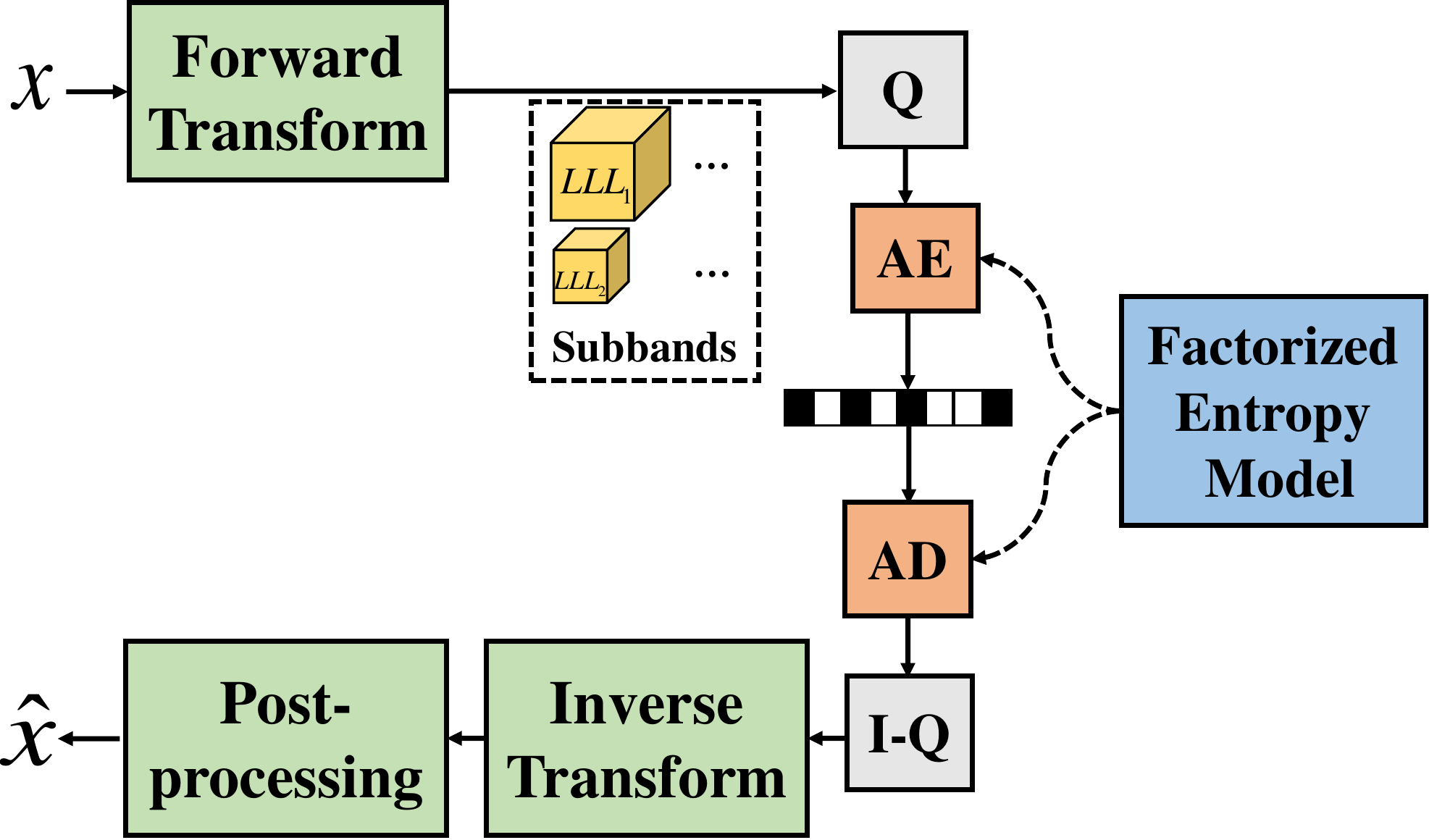}}
\caption{Overview of the proposed iWave3D. ``Q" and ``I-Q" stand for quantization and inverse-quantization, ``AE" and ``AD" stand for arithmetic encoding and arithmetic decoding respectively. The forward transform and inverse transform modules are implemented with 3-D lifting scheme (see Fig. \ref{fig-lifting} for the description of 1-D lifting scheme), in which the predict and update steps are implemented with 3-D convolutional networks (see Fig. \ref{fig-network structure}). Note that the forward and inverse transform modules share the same set of parameters. A simple factorized model is utilized for entropy coding. A post-processing module is utilized to compensate for quantization error. iWave3D could support lossless compression by discarding the quantization, inverse quantization, and post-processing modules, since the transform module does not introduce any information loss.}
\label{fig-iWave3D}
\end{figure}

In this section, we introduce the details of the proposed iWave3D. Firstly, we give the overall structure of the end-to-end compression network as shown in Fig. \ref{fig-iWave3D}. After that, we introduce the detail of transform module which is the highlight of our scheme. Then the post-processing module is described. Finally, we give the loss function to train the whole model.
\subsection{Overview of iWave3D}
The iWave3D consists of wavelet forward transform, quantization, entropy coding, inverse quantization, wavelet inverse transform, and post-processing.
At encoder part, the wavelet forward transform $g_a(;\phi)$ converts $x$ into wavelet coefficients $y$, $y=g_a(x;\phi)$, where $g_a(;\phi)$ is implemented with 3-D trainable lifting scheme, which will be described in Section \ref{transform}.
Then $y$ is quantized and rounded to the nearest integer, $q =[y/QS]$, where [$\cdot$] is the rounding operation and $QS$ is the quantization step. In our experiment, $QS$ is set as a trainable variable, and will be determined after training. Finally, the factorized entropy model\cite{balle2016end,balle2018variational} assists the arithmetic encoder to write $q$ into bit stream. The trainable parameters in the entropy model will also be determined after training.

\begin{figure}
\centering
\includegraphics[width=0.48\textwidth]{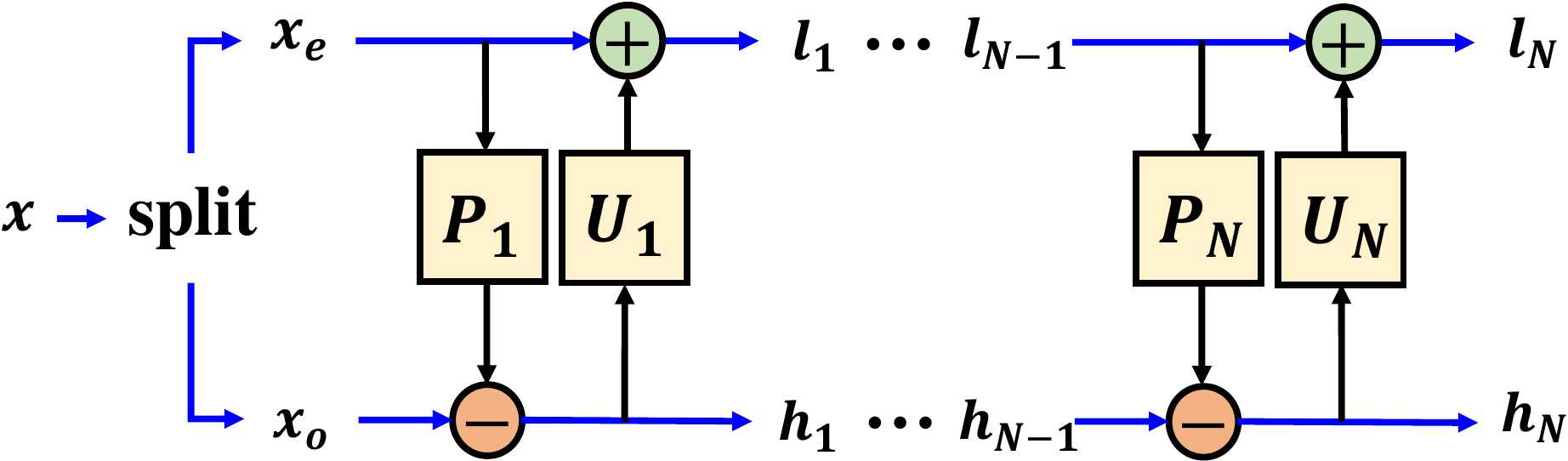} \\
\includegraphics[width=0.48\textwidth]{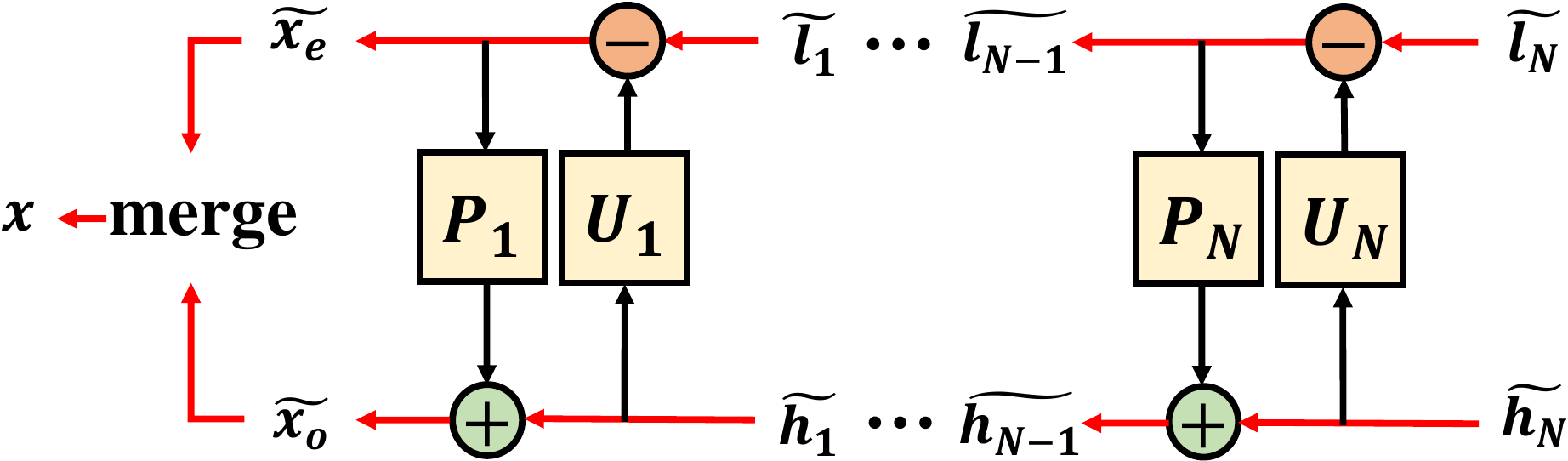} \\
\caption{Forward (above) and inverse (below) transform of trainable lifting scheme for 1-D signal. ``P'' stands for predict network and ``U'' stands for update network. We use $N=2$ in this paper. After lifting the input signal $x$, a low frequency subband $l_N$ and a high frequency subband $h_N$ are obtained. It can be seen that by using the same set of parameters in forward and inverse transform, and letting $\widetilde{l_N}=l_N$ and $\widetilde{h_N}=h_N$, then $\widetilde{x}=x$. This indicates that the proposed transform is revertible.}
\label{fig-lifting}
\end{figure}

At decoder part, the bit stream is first decoded to get $q$. Then the inverse-quantization is followed to reconstruct $\hat{y}$, $\hat{y}=q\times QS$.
At last, the restored image $\hat{x}$ is obtained after wavelet inverse transform, $\hat{x}=g_a^{-1}(\hat{y};\phi)$. Note that wavelet inverse transform $g_a^{-1}$ is the inverse of wavelet forward transform $g_a$, and they share the same set of parameters $\phi$.
The post-processing module is used to compensate for quantization loss and enhance the restored image.

It is worth noting that the transform module we propose is reversible, and the information loss only occurs in the quantization process. As a result, by simply removing the quantization, inverse quantization, and post-processing steps, iWave3D could be used for lossless compression. The performance of iWave3D for lossy compression and lossless compression is shown and analysed in section \ref{3}. 

\subsection{3-D trainable Wavelet Transform}\label{transform}
To implement a trainable wavelet transform, we take advantage of the lifting scheme, which is also called lazy wavelet transform. The basic idea of lifting is to cascade simple predict and update filters to form a complex wavelet transform.

We first take the 1-D lifting as example to show the lifting process. As shown in Fig. \ref{fig-lifting}, one lifting step is mainly divided into three steps: split, predict and update. The input signal $x$ is first \textit{split} into an odd part $x_o$ and an even part $x_e$. 
Then by using appropriate filters, the \textit{predict} $P$ and the \textit{update} $U$ are performed alternately to obtain the high frequency component $h$ and the low frequency component $l$. 
Note that the inverse transform is the inverse process of the above steps.

The 3-D lifting is to successively perform 1-D lifting in the axial, horizontal and vertical directions. For a 3-D image, after lifting in the axial direction, we get two subbands $L$ and $H$. Then by performing lifting in the horizontal direction, $L$ is transformed into $LL$ and $HL$, and $H$ is transformed into $LH$ and $HH$. For the resulting four subbands, finally, another lifting is carried out in the vertical direction, and eight subbands \{$LLL$, $HLL$, $LHL$, $HHL$, $LLH$, $HLH$, $LHH$, $HHH$\} are obtained. 
Usually the subband $LLL$ continues to be decomposed to form a pyramid structure. Through experiments, we find that when the decomposition level is greater than 2, the performance is improved only slightly. As a result, we fix the decomposition level to 2 in this paper.

\begin{figure}[t]
\centerline{\includegraphics[width=0.98\columnwidth]{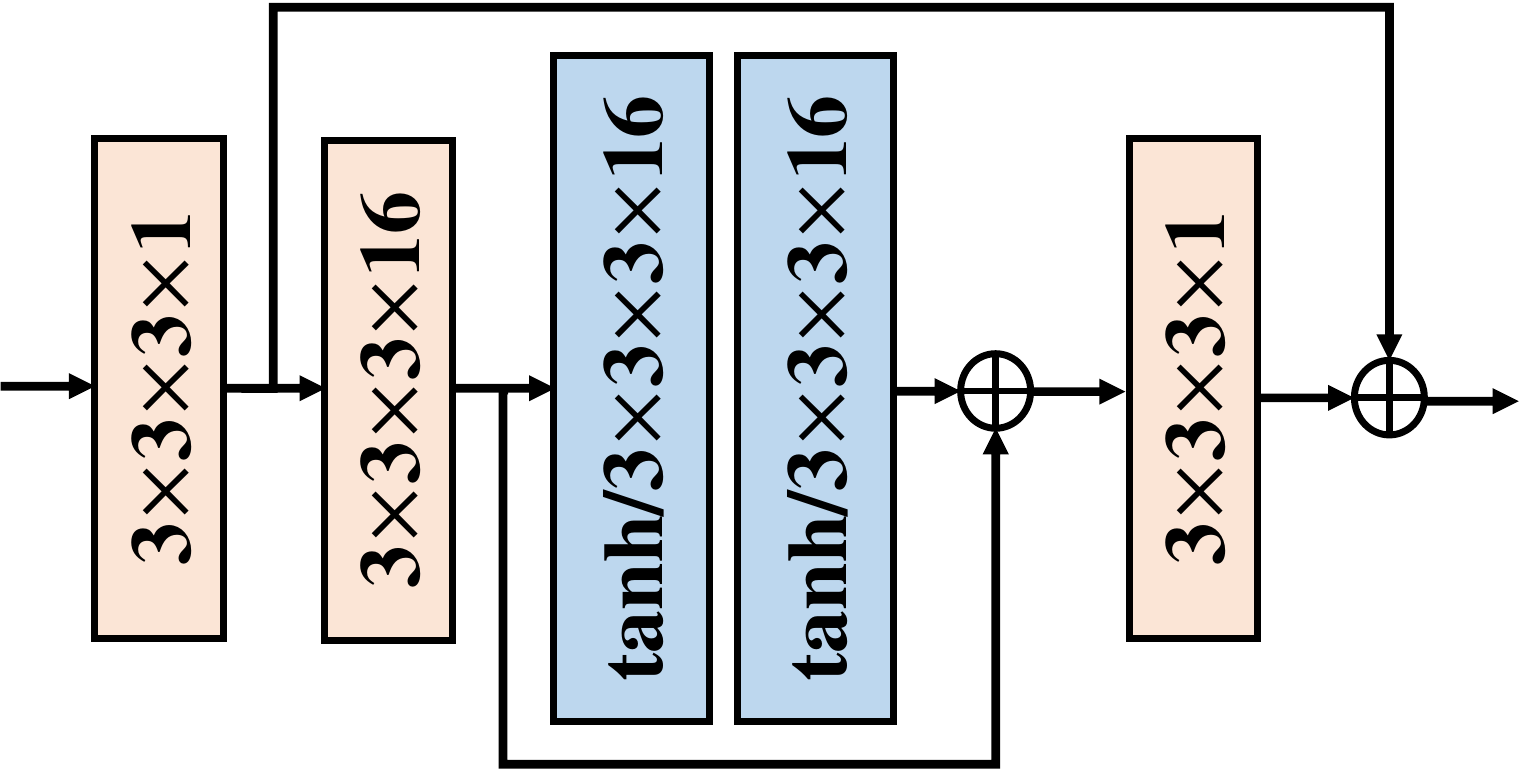}}
\caption{3-D convolutional network used for each predict and update steps (see Fig. \ref{fig-lifting}). The shown numbers like $3\times3\times3\times16$ indicate the kernel size ($3\times3\times3$) and the number of channels (16) of each layer. ``tanh'' indicates the adopted nonlinear activation function.}
\label{fig-network structure}
\end{figure}

In this paper, the predict and update steps in the above 3-D lifting process are all implemented with 3-D CNNs, resulting in a trainable 3-D wavelet transform. The structure of the used 3-D CNN is shown in Fig. \ref{fig-network structure}. It is worth noting that we have shared the parameters of CNNs in different lifting steps to reduce complexity. By training with brain images, our 3-D trainable wavelet transform is much more efficient than traditional wavelet transform when compressing brain images.

\subsection{Post-processing }
In iWave3D, we introduce a post-processing module to compensate for quantization loss. Since our transform is reversible, the loss caused by quantization cannot be compensated by the inverse transform. This is quiet different from other popular end-to-end compression frameworks.

\subsection{Loss Function}
iWave3D is directly trained to minimize the rate-distortion loss.
\begin{equation}
L = \mathbb{E}{_{x \sim {D_x}}}\left[ { - {{\log }_2}{p_{q}}\left( {q} \right)} \right] + \lambda  \cdot  {\mathbb{E}{_{x \sim {D_x}}}\left\| {x - \hat x} \right\|_2^2}
\end{equation}
where $D_x$ stands for the training dataset, $x$ stands for a training sample from $D_x$, $\hat{x}$ stands for the reconstruction, and $q$ stands for the quantized coefficients. Here we use mean square error to calculate the reconstruction loss. $\lambda$ is the Lagrange multiplier to balance the two loss terms. 

\section{Experiment Results}
\label{3}
\subsection{Simulation Setup}
We train our iWave3D on FAFB dataset, which is collected by serial section Transmission EM (ssTEM)\cite{knott2008serial} technology. The resolution of FAFB dataset is 4 × 4 × 40 \(nm\). Note that the resolution in the axial direction is much lower than that in the horizontal and vertical directions due to technical limitations. We use 1124 cubes totally, of which 1024 cubes are for training and 100 cubes are for testing. The cube size is $64\times64\times64$. The Adam optimizer is utilized for training. During training, the learning rate is set to $10^{-4}$.

We evaluate our iWave3D on test data with peak signal-to-noise ratio (PSNR) and structural similarity (SSIM) \cite{wang2004image}. The compared methods include HEVC and JP3D. 
For HEVC, we use the reference software HM-16.15\footnote{https://vcgit.hhi.fraunhofer.de/jvet/HM/-/tree/HM-16.15} with RA configuration.
As for JP3D, we use OpenJPEG 2.3.1\footnote{http://www.openjpeg.org/2019/04/02/OpenJPEG-2.3.1-released} with default configuration.
\begin{table}[t]
\centering
\caption{Lossless Compression Performance on FAFB test set.}
\fontsize{9.0}{8}\selectfont
\begin{tabular}{c|c|c|c}
\hline
Methods &JP3D-LS & HEVC-LS & iWave3D-LS \\ \hline
Rate  & 5.185 & 6.218  & 5.609 \\ \hline
\end{tabular}
\label{lossless performance3}
\end{table}

\subsection{Performance}
\begin{figure*}[h]
\centering	
\subfigure
{
\includegraphics[width=0.92\columnwidth]{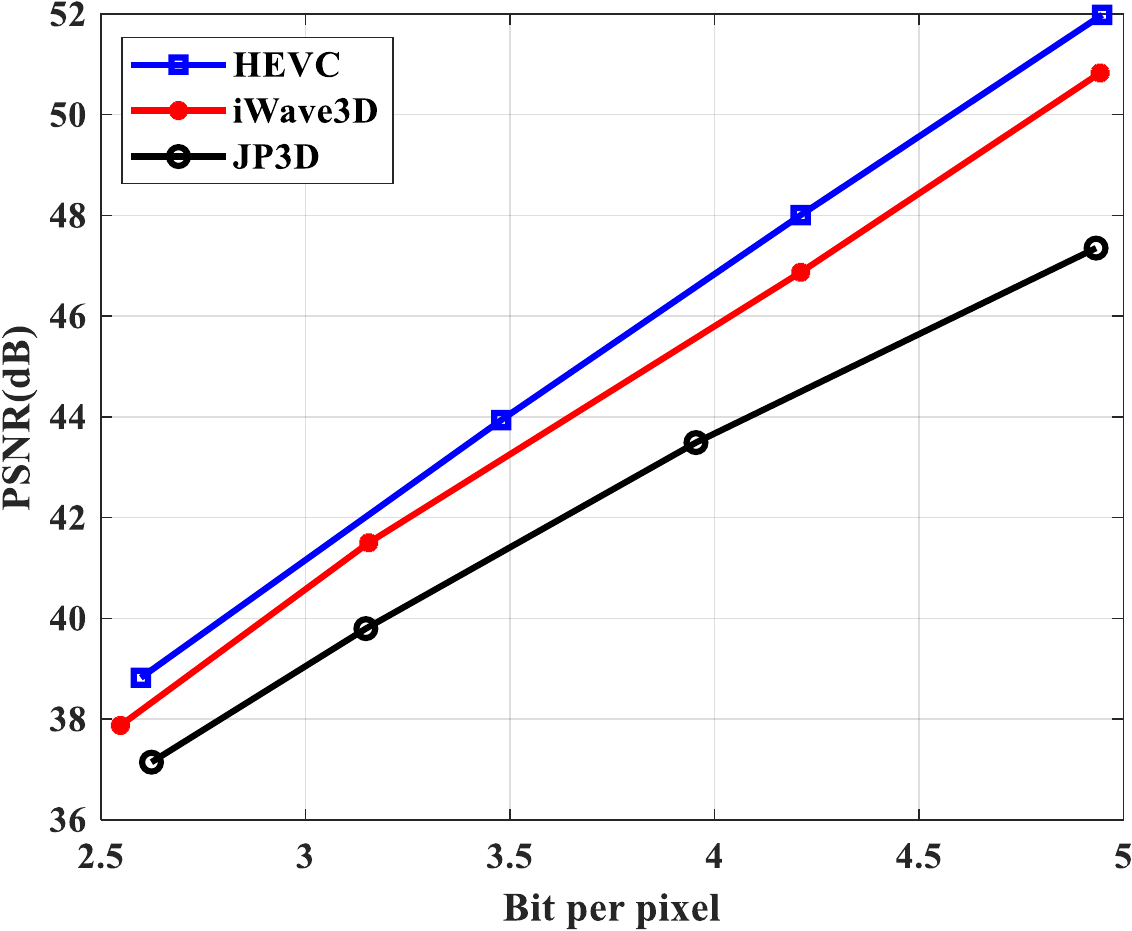}
}
\subfigure
{
\includegraphics[width=0.94\columnwidth]{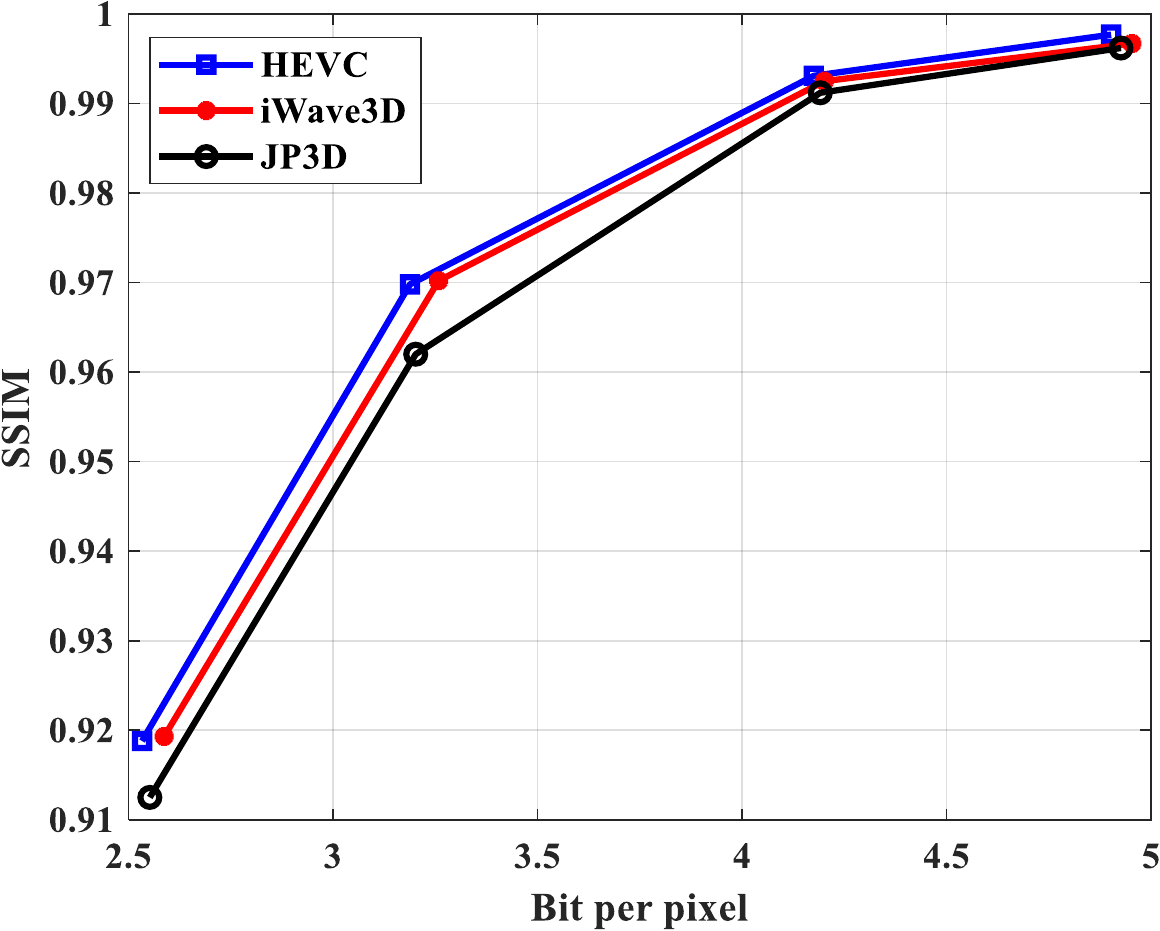}
}
\caption{Comparison of different lossy compression methods on the
FAFB test set. The figure on the left shows the average rate (bits-per-pixel) and the average PSNR of the 100 test brain images. The figure on the right shows the average rate and the average SSIM
of the 100 test brain images.}
\label{fig-PSNR curve}
\end{figure*}

\begin{figure*}[h]
\centering	\subfigure
{
\includegraphics[width=0.865\columnwidth]{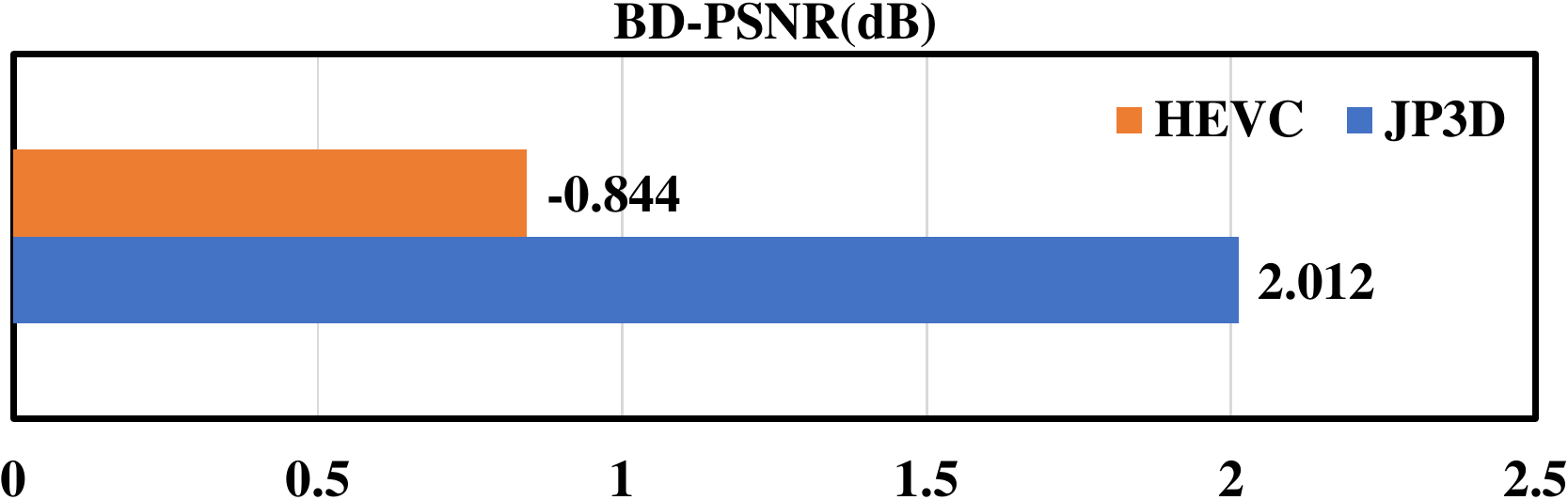}
}
\subfigure
{
\includegraphics[width=0.88\columnwidth]{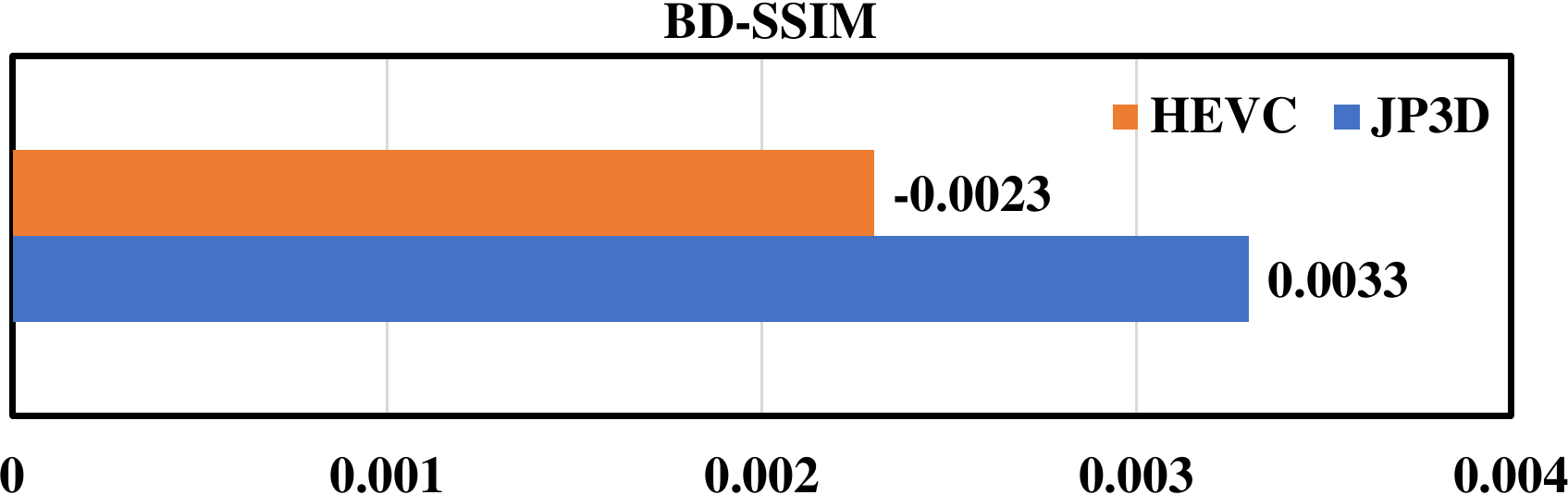}
}
\caption{Shown are the average BD-PSNR and BD-SSIM on FAFB test set of iWave3D compared with JP3D and HEVC for lossy compression.}
\label{fig-BDPSNR}
\end{figure*}

Fig.~\ref{fig-PSNR curve} shows the rate-distortion curves of the proposed iWave3D, JP3D, and HEVC for lossy compression. 
Fig.~\ref{fig-BDPSNR} shows the average Bjontegaard Delta-PSNR (BD-PSNR) and Bjontegaard Delta-SSIM (BD-SSIM), respectively. 
 Note that our method is only optimized for mean square error(MSE) rather than SSIM.
It can be seen that our iWave3D outperforms JP3D for 2.012 dB on average. The SSIM result of iWave3D also outperforms JP3D slightly. Compared with HEVC, our iWave3D performs worse. This is mainly because iWave3D is a lightweight codec, it only uses a simple entropy coding module, but the complexity of HEVC is very high.

\begin{figure*}[h!]
\centerline{\includegraphics[width=1.95\columnwidth]{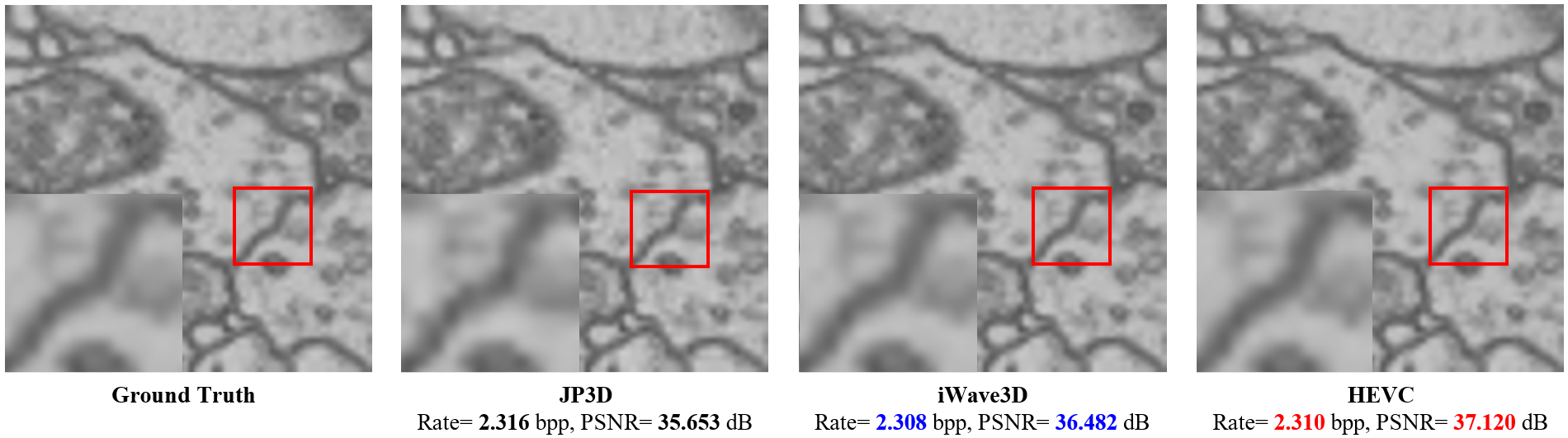}}
\caption{Visual examples from FAFB test set of JP3D, HEVC and iWave3D (ours).}
\label{fig-vis}
\end{figure*}

For lossless compression, the comparison results of iWave3D, JP3D and HEVC are given in Table \ref{lossless performance3}. JP3D and HEVC are also tested on their lossless compression mode. Our iWave3D saves 0.237 bpp on average compared with JP3D.

\subsection{Visualization}
Fig. \ref{fig-vis} shows visual examples of the three methods. Obviously it can be seen that our iWave3D has better visual quality than JP3D. In particular, the edges are sharper and there are fewer artifacts. The visual results of HEVC and iWave3D are also comparable.

\section*{Conclusion}

In this paper, an end-to-end framework iWave3D is proposed for brain image compression. The highlight module of our method is a 3-D trainable wavelet transform. We solve the shortcomings of traditional wavelet transforms that are designed manually for ideal signals and cannot be optimized for specific dataset. Moreover, the transform we proposed is reversible, thus our iWave3D can be utilized for both lossy and lossless compression to satisfy the requirement of high-quality reconstruction of brain images. By adding a simple factorized entropy module and a post-processing module, we achieve an end-to-end compression framework. All the modules are jointly optimized by minimizing the rate-distortion loss function. The experimental results show the superiority of our iWave3D.

\section{Acknowledgement}
This work was supported by USTC Research Funds of the Double First-Class Initiative Grant YD3490002001 and The University Synergy Innovation Program of Anhui Province No. GXXT-2019-025.
It was also supported by the GPU cluster built by MCC Lab of Information Science and Technology Institution, USTC.

\bibliographystyle{IEEEtran}
\bibliography{refer}

\end{document}